\documentclass{article}

\usepackage[preprint]{neurips_2022}

\usepackage[utf8]{inputenc} 
\usepackage[T1]{fontenc}    
\usepackage{hyperref}       
\usepackage{url}            
\usepackage{booktabs}       
\usepackage{amsfonts}       
\usepackage{nicefrac}       
\usepackage{microtype}      
\usepackage{xcolor}         
\usepackage{graphicx}
\usepackage{subfigure}
\usepackage{booktabs}
\usepackage{parskip}
\usepackage{amsmath}
\usepackage{amssymb}
\usepackage{mathtools}
\usepackage{amsthm}
\usepackage{tabularx}
\title{Capturing cross-session neural population variability through self-supervised identification of consistent neuron ensembles}

\author{Justin Jude\\
University of Edinburgh\\
\texttt{justin.jude@ed.ac.uk}
\And
Matthew G. Perich \\
Icahn School of Medicine at Mount Sinai \\
New York, NY 10029 \\
\texttt{mperich@gmail.com} \\
\And
Lee E. Miller \\
Feinberg School of Medicine \\
Northwestern \\
Chicago, IL 60611\\
\texttt{lm@northwestern.edu} \\
\And
Matthias H. Hennig\\
University of Edinburgh\\
\texttt{m.hennig@ed.ac.uk}
}
\newcommand{\argmax}{\operatornamewithlimits{argmax}}

\begin{document}

\maketitle

\begin{abstract}
Decoding stimuli or behaviour from recorded neural activity is a common approach to interrogate brain function in research, and an essential part of brain-computer and brain-machine interfaces. Reliable decoding even from small neural populations is possible because high dimensional neural population activity typically occupies low dimensional manifolds that are discoverable with suitable latent variable models. Over time however, drifts in activity of individual neurons and instabilities in neural recording devices can be substantial, making stable decoding over days and weeks impractical. While this drift cannot be predicted on an individual neuron level, population level variations over consecutive recording sessions such as differing sets of neurons and varying permutations of consistent neurons in recorded data may be learnable when the underlying manifold is stable over time. Classification of consistent versus unfamiliar neurons across sessions and accounting for deviations in the order of consistent recording neurons in recording datasets over sessions of recordings may then maintain decoding performance. In this work we show that self-supervised training of a deep neural network can be used to compensate for this inter-session variability. As a result, a sequential autoencoding model can maintain state-of-the-art behaviour decoding performance for completely unseen recording sessions several days into the future. Our approach only requires a single recording session for training the model, and is a step towards reliable, recalibration-free brain computer interfaces.
\end{abstract}

\section{Introduction}

Neural decoders require stable neurons in a recorded population in order to accurately predict behaviour such as movement or to allow decoding of stimuli. However, over time instabilities in the recording equipment and drift in neural activity lead to instabilities that prevent re-using a decoder trained on one day for a session recorded on another day \citep{huber2012multiple,Ziv2013,driscoll2017dynamic}. At the same time, neural population activity is highly structured and often confined to low-dimensional manifolds \citep{cunningham2014dimensionality} that can be recovered using latent variable modelling approaches \citep{Hurwitz2021BuildingPitfalls}. Importantly, recent work showed that movement-related latent neural dynamics in population activity from the primate motor cortex is stable and could be recovered over intervals as long as two years \citep{Gallego2020Long-termBehavior}. This suggests that despite the variability at the level of single neurons, in each session a subset of neurons will remain informative about behaviour. A stable cross-session decoder therefore has to be able to identify these neurons and utilise them for decoding. Therefore, here we focus on identifying known recording neurons in unseen sessions. In particular, we hypothesised that a latent encoding of neural activity can be augmented by information about which neurons were seen during training, and at which position in the input. We show that this is sufficient to decode behaviour (in our case different cued arm movements by a monkey with simultaneous motor cortex recordings) with high accuracy across unseen sessions.

\begin{figure}[h!]
\begin{center}
\includegraphics[width=0.9\textwidth]{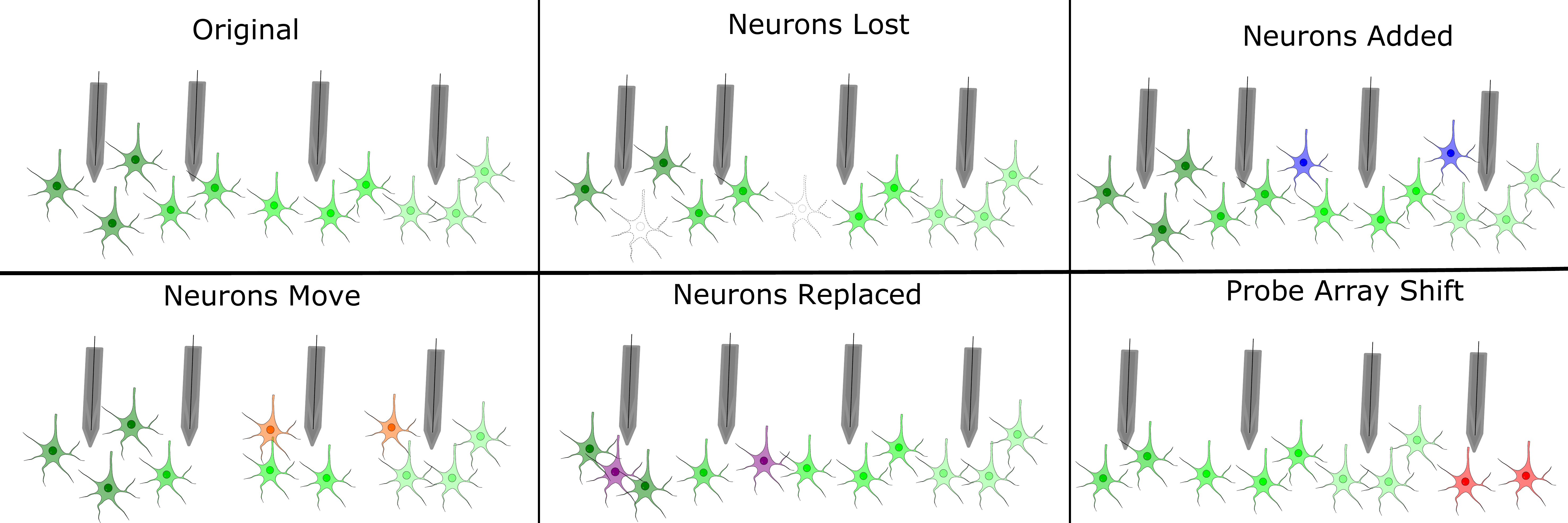}
\caption[variability]{Inter-session ensemble variability possible when recording from neural populations. Neurons from the original recording session can be lost to the recording array, new neurons can become visible, neurons can move between electrodes, original neurons can be replaced by unseen neurons and the entire probe array can shift, causing a systematic change in neuron position. In addition, spike sorting can induce variability as the signal to noise ratio of individual neurons changes between sessions. The perturbations we apply to each trial of recordings is in response to each of these sources of variability. We model each unseen test trial as an instance of a perturbed seen train trial and subsequently, our sequential autoencoder model attempts to map each unseen trial to a known trial.}
\label{fig:variability}
\end{center}
\end{figure}

We achieve this with a self-supervised approach through training a recurrent neural network (RNN) to predict original neuron positions following data perturbation in a manner mirroring session to session variability. In essence, the closer our perturbations mimic real inter-session variability (as shown in Figure \ref{fig:variability}), the higher our behaviour prediction performance on an unseen session. These perturbations include adding spikes to existing neurons from randomly generated neurons, removing spikes from existing neurons, shifting the entire neuron population by a constant amount, slightly shifting neurons in time, replacing neurons with randomly generated neurons and eliminating neurons entirely. 

This neuron locator RNN is trained to predict original neuron position within a single recording session from many perturbed variations of trials of this training session. Once trained to predict original neuron positions, a separate network, which in this case is a sequential autoencoder based on Latent Factor Analysis via Dynamical Systems (LFADS) \citep{Pandarinath2017InferringAuto-encoders}, is trained to predict original unperturbed neural recording trials from perturbed variations of trials from the same session. The encoder of this sequential autoencoder receives as additional input the embedding of the neuron locator RNN activations, conditioning the encoder to produce latent variables which are informative enough to accurately reconstruct the original recording. The encoder produces latent variables which are separated by behaviour (arm movement direction) in a self-supervised manner, from which behaviour can be predicted without the model being explicitly trained on behaviour.

Importantly, the joint neuron locator RNN and LFADS encoder ensemble can predict behaviourally relevant latent variables for unseen recording sessions that yield high decoding accuracy. Currently, there are no existing approaches to accurately predict behaviour from an unseen recording session when training on just one single session. We not only show this is possible with our method, but that our approach is robust to inter-session variability for up to 8 days when a sufficient number of neurons are persistent across sessions.

\section{Related Work}
There have been many recent approaches to creating robust behaviour decoders of neural activity
\citep{Gallego2020Long-termBehavior,Farshchian2019AdversarialInterfaces,Sussillo2016MakingVariability,Wen2021RapidModelling,Karpowicz2022.04.06.487388,Wimalasena2021.12.01.470827}. However these methods are not capable of decoding behaviour from a previously unseen recording session if the recorded activity is subject to random fluctuations.

Recent work in modelling neural activity shows the consequences of selectively perturbing neural data in order to learn relevant latent variables in a self-supervised way using an autoencoder \citep{liu2021drop,https://doi.org/10.48550/arxiv.2102.10106,NEURIPS2021_1325cdae}. These models take different views of the same neural data and align the latent spaces of these views once passed through an encoder, with the ultimate aim of reconstructing these views. We utilise a similar technique to train our sequential autoencoder by aligning the latent variables of perturbed versions of the same data and aim to generate the activity of the original unperturbed trial. Importantly, \cite{liu2021drop} propose a model which is invariant to the specific neurons used to represent the neural state within training data; in this work we look at unseen sessions and so do not aim to produce a model invariant to new neurons, but one that is able to identify and utilise seen neurons to reconstruct unperturbed trials.

\citet{gonschorek2021removing} and \cite{https://doi.org/10.48550/arxiv.2202.06159} use domain adaptation to align data across recording sessions. In both studies the authors use an autoencoder model and a domain classifier. However these models require training on many days of recording sessions for good behaviour decoding accuracy. For instance, \cite{https://doi.org/10.48550/arxiv.2202.06159} requires as many as 12 training sessions and training on behaviour explicitly in order to produce high behaviour decoding accuracy on an unseen test session. In this work we achieve state-of-the-art behaviour decoding performance on an unseen test recording session using just one training recording session, and show that this decoding accuracy can be maintained many days into the future without recalibration.

We train an RNN to predict original neuron position from perturbed trials and utilise this network to inform the sequential autoencoder model. This is considered self-supervised learning as we do not train our model on behaviour explicitly but instead train on the subtasks of predicting original neuron positions and reconstructing unperturbed trials from perturbed ones. This approach is similar to that used in \cite{Noroozi2016UnsupervisedPuzzles}, where authors form 9 subsets of images and randomly permute these subsets, then task the model with predicting the permutation. 


\section{M1 Recordings}

\begin{figure}[h!]
\begin{center}
\includegraphics[width=0.3\textwidth]{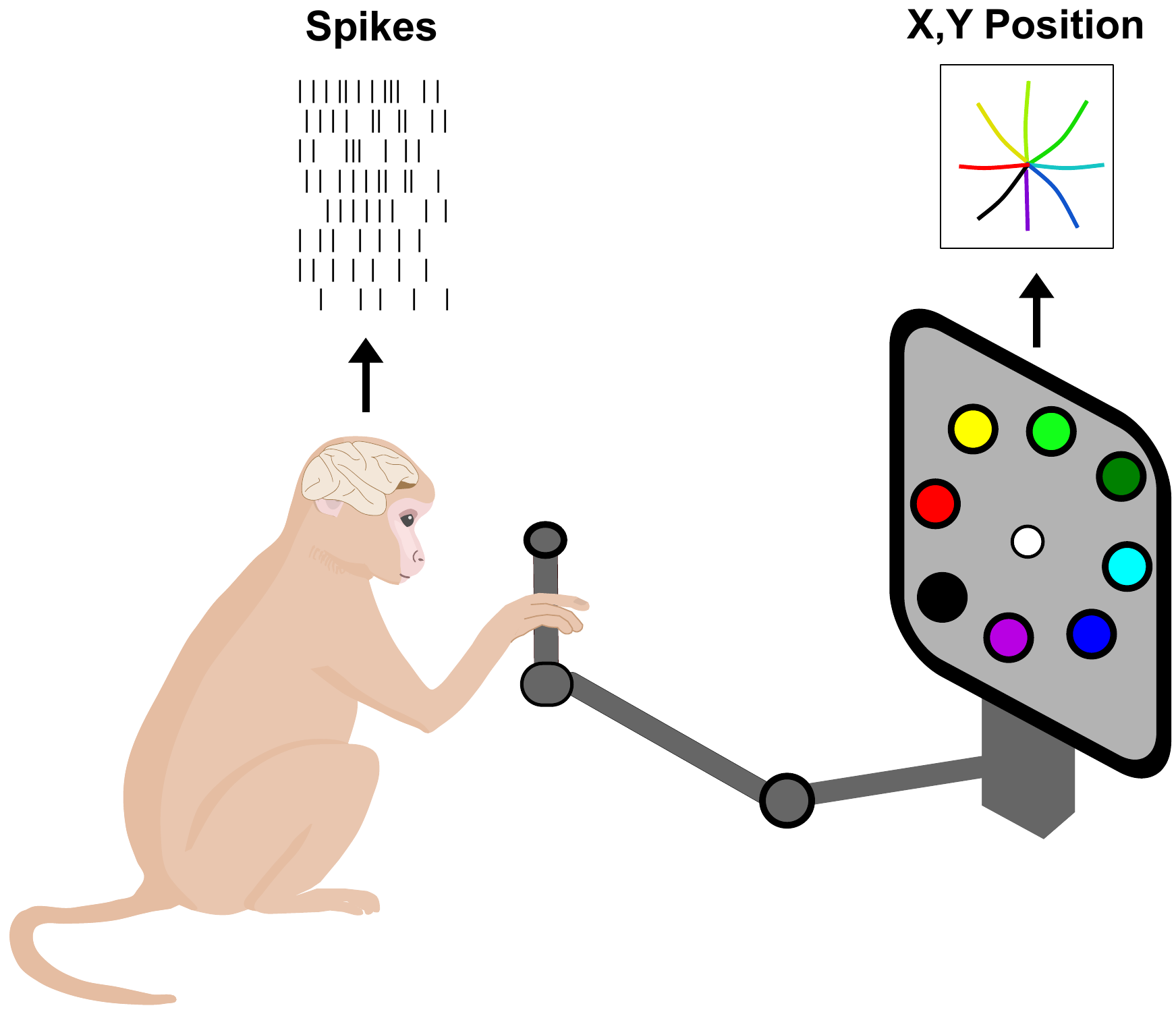}
\caption[experimentsetup]{Experimental setup:  In each trial one randomly chosen target direction (indicated by one of 8 coloured circles) appears on screen, and the monkey is instructed to control the cursor (white circle) by moving the manipulandum. The monkey moves the cursor to the target location after a go cue. The collected data for each trial consists of the neural spikes and monkey hand position across all timesteps. We predict hand position from neural spikes at each timestep.}
\label{fig:experiment}
\end{center}
\end{figure}

We apply our model to data from a previously published experiment \citep{Gallego2020Long-termBehavior}. In this experiment, two monkeys were trained to perform a center-out reach task towards eight outer targets. On a go cue, each monkey moves a manipulandum along a 2D plane to guide a cursor on a screen to the target location (Figure \ref{fig:experiment}). On successful trials a liquid reward is given. Spiking activity from the motor cortex (M1) along with the 2D hand position were recorded during each trial. Spike trains were converted into spike counts in 10ms bins, and behaviour variables are used at the same resolution. In this work, only successful trials are used, all trials are aligned to movement onset and cut from movement onset to the shortest reach time across all trials.

For our analysis, we train our model on one session of recorded data from a single day which we denote day 0 (containing 173 trials for both monkeys) and test on subsequent held out days of recordings for each monkey. A comparison of the activity between sessions shows considerable variability, caused by shifts in the order neurons appear as well as disappearance of neurons and the appearance of new ones (see Appendix B, Figure 8). These changes are particularly pronounced for longer time intervals, but are already significant in recordings one day apart. In total we used 5 days of recordings for both monkeys, with 55 recorded neurons across all sessions for Monkey C and 17 for Monkey M. Each day for each monkey consists of one recording session.

\section{Data Perturbations}
\begin{figure}[h!]
\begin{center}
\includegraphics[width=0.95\textwidth]{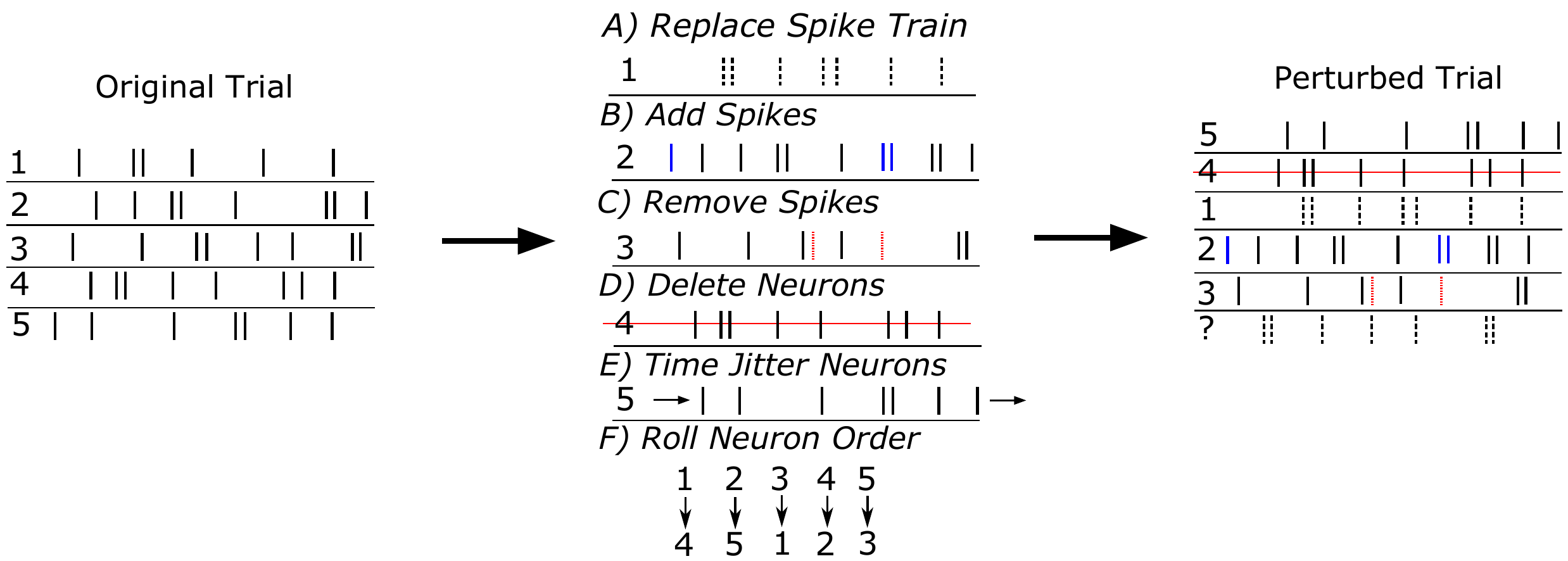}
\caption[perturbations]{Perturbations applied simultaneously to each trial of neural data, demonstrated with a simple 5 neuron system. A) Replace entire spike train with a randomly generated neuron of the same firing rate as the original neuron. B) Spikes randomly added to spike train proportional to average firing rate of all neurons in a given trial, to mirror influence of nearby newly added unknown neurons. C) Spikes randomly removed to mimic removal or movement of nearby known neurons. D) Deletion of entire neurons to simulate neuron loss between sessions, with randomly generated neurons introduced as the first or last neuron of the trial to keep neuron number consistent. E) Small random time jitter of all neuron spike trains to simulate experimental variation between sessions. F) Constant random shift of the order of all neurons to mirror probe shift.}
\label{fig:perturbations}
\end{center}
\end{figure}

Fig. \ref{fig:perturbations} outlines the perturbations forming each variation of a single trial during the training of our model. Perturbations A) to D) in Fig. \ref{fig:perturbations} are applied with equal probability to a given neuron of a given trial. Perturbation E) is applied to all neurons, time jitter is chosen randomly between -30ms and +30ms. Perturbation F) is applied to all trials, the amount of this neuron shift is chosen randomly between 0 and 25\% of the total number of neurons. We hypothesise that this combination of transformations sufficiently mirrors the real day to day changes of recorded neuron ensembles.

\section{Model}
\begin{figure*}[ht!]
\begin{center}
\includegraphics[width=0.62\textwidth]{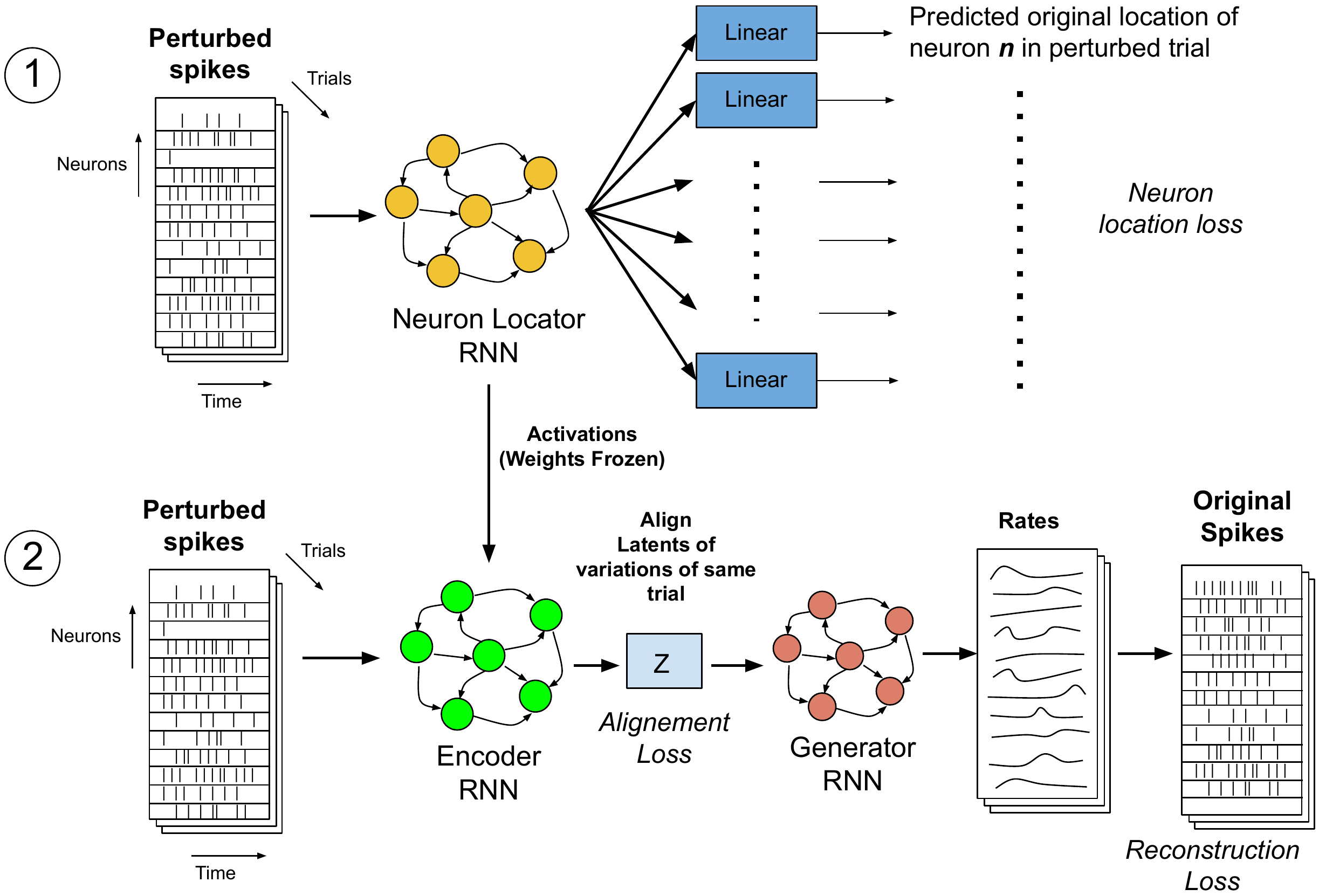}
\caption[lfadsDA]{Our model consists of a neuron locator RNN (1) combined with a sequential variational autoencoding approach (2). The neuron locator (1) is trained first to identify original neuron position (or if the neuron is randomly generated) in each trial after perturbations have been applied. Then the neuron locator's weights are frozen and its activations are given as additional input to condition the encoder of the sequential autoencoder (2). Notably, we perturb recording trials when training both the neuron locator and sequential variational autoencoder. The sequential autoencoder is tasked with reconstructing the original unperturbed recording trials. The encoder of the sequential autoencoder maps perturbed versions of the same trial to similar latent variables. This is accelerated by imposing an alignment loss across the latent variables of variations of the same trial. The generator RNN of the sequential autoencoder predicts original trials from latent variables produced by the encoder RNN.}
\label{fig:lfadsDAmodel}
\end{center}
\end{figure*}

Our modelling approach is based on the hypothesis that the perturbations mentioned above can capture the substantial variability between recording sessions from the same animal. We also expect neural activity $x$ is related to the latent variables $z$ through a simple function, however, this function will differ between recording sessions as we expect to observe different neurons in each session. The problem is thus to find the correct encoding function $z=f(x)$ to transform perturbed neural activity into a consistent latent space which then allows decoding of behaviour. In addition, for the same behaviour we require $z_i$ for each trial $i$ to be similar despite variations in the activity $x_i$. 

We first train a fully connected layer and an RNN to predict original neuron position in perturbed trials. We apply the perturbations from Figure \ref{fig:perturbations} to each trial, then task the network to predict the original position of each neuron in the recording data or whether it was previously unseen. As shown in Figure \ref{fig:lfadsDAmodel}, for each neuron in the recording data we project a softmax linear read-out layer from the RNN which each form a probability distribution of predicted original neuron position across all possible positions (plus an extra position indicating that the neuron was randomly generated). Each of these is compared against a one hot encoding of the original neuron position before any perturbations have been applied. If the neuron is randomly generated then the one-hot encoding is one at the dedicated extra position. Predictions of original neuron position are made as follows:
\begin{gather} \label{eq:jigsaw}
\bar{x}_{i,1:T} = \text{Perturb}(x_{i,1:T}), \\
\text{acts}_i = \text{GRU}_{\theta_{\textrm{pos}}}(f_{\textrm{pos}}(\bar{x}_{i,1:T})),\\ 
\text{pos}_{i,n} =  \text{softmax}(W^n_{\textrm{neuron}} . \text{acts}_i) \\ \nonumber
\text{The predicted position for trial $i$ and neuron $n$ is then:} \argmax \text{pos}_{i,n}
\end{gather}
Perturb is the simultaneous application of all perturbations outlined in Section 4 to a given trial. $f_{\textrm{pos}}$ is a fully connected layer and $\theta_{\textrm{pos}}$ are the parameters of the locator network used to predict original neuron position. $W^n_{\textrm{neuron}}$ is the set of linear layers used to predict original neuron position, producing a probability distribution when combined with a softmax layer for each neuron.

Once trained, the weights of this neuron locator network are frozen, and the activations of the RNN are used as additional input to the encoder of an LFADS-inspired sequential autoencoder. This input conditions the encoder in predicting latent variables used to generate original trials from perturbed trials. As proposed by \cite{Pandarinath2017InferringAuto-encoders} we assume that the latent dynamics evolve autonomously provided a set of initial conditions $z_i$ that are modelled as Gaussian random variables. These latent variables are produced for each trial by an encoder network consisting of bidirectional Gated Recurrent Units \citep{Cho2014LearningTranslation} (GRU). They are used to reconstruct the original trial-specific neural activity from the perturbed trials. A further bidirectional GRU is used as a generator for neural reconstruction of unperturbed trials from latent variables $z_i$. Training is based on Poisson likelihood for unperturbed neural activity reconstruction (as in \citep{Pandarinath2017InferringAuto-encoders}). The model is trained using real neural activity which corresponds to consistent behaviours (movement directions in a centre-out reach task). The generative process of our model is as follows:
\begin{gather} \label{eq:gen}
z_i = f_{\textrm{enc}}(\text{GRU}_{\theta_{\textrm{enc}}}(\bar{x}_{i,1:T}; \text{acts}_i)),\\g_{1:T} =  \text{GRU}_{\theta_{\textrm{gen}}}(z_i), \\
r_t =  exp(W_{\textrm{rate}} . f_{\textrm{fac}}(g_t)),\\
\hat{x}_t \sim \text{Poisson}(r_t)
\end{gather}

where $\theta_{\textrm{enc}}$ and $\theta_{\textrm{gen}}$ are the parameters of the GRUs used to encode perturbed spike trains into latent variables and subsequently generate original unperturbed spike trains from the latent variables. $f_{\textrm{enc}}$ and $f_{\textrm{fac}}$ are fully connected layers which produce latent variables and neural activity factors respectively. $W_{\textrm{rate}}$ is a linear transformation used to generate firing rates at each time step per trial. At each training iteration the following three losses are optimised with Adam \citep{Kingma2015Adam:Optimization}:
\begin{gather}
L_{{\textrm{rec}}} = -\sum_{t=1}^{t} \log(\text{Poisson}(x_{i,t}\vert r_t)) \label{eq:rec} \\
L_{kl} = D_{KL}[\text{GRU}_{\theta_{\textrm{enc}}}(z_i \vert \bar{x}_i; \text{acts}_i) \vert \vert \mathcal{N}(0, I)]
 = - \frac{1}{2}[\log(z_{i,\sigma}^2) - z_{i,\mu}^2 - z_{i,\sigma}^2 + 1] \label{eq:kl}  \\
 L_{\textrm{align}} = \frac{1}{P} \sum_{j=1}^{p} \sum_{k\neq j}^{p}(z_{i,j} - z_{i,k})^2 \label{eq:align} 
\end{gather}

Together $L_{\textrm{rec}}$ and $L_{kl}$ are the usual evidence lower-bound
of the marginal log-likelihood in a VAE \citep{Kingma2014Auto-encodingBayes}. $L_{\textrm{rec}}$ is minimised by the encoder network and the neural generator network. As in \cite{liu2021drop}, we apply an alignment loss ($L_{\textrm{align}}$) across latent variables produced from perturbed trials (where $P$ is the number of perturbations of a given trial) of the same original trial $z_i$ which reduces training duration. We form 2 perturbed variations of each trial in a given batch at each training iteration. Kullback–Leibler ($L_{kl}$) divergence loss (between a multivariate standard Gaussian distribution and the encoder-generated latent variables) and $L_{align}$ are minimised by just the encoder network. We name our model CAPTure and Identify Variability at Target Ensembles (CAPTIVATE). Further implementation details can be found in Appendix A.

\subsection{Comparison models}

We compare the ability of CAPTIVATE to predict behaviour from sessions of unseen spike data against existing methods and against a variation of our own model where we do not use the locator network trained on original neuron position to aid in aligning perturbed trials. We denote this model variation CAPTIVATE-noLoc. In addition, we look at vanilla LFADS \citep{Pandarinath2017InferringAuto-encoders} in  autoencoding trials without any perturbations. We also compare against a baseline RNN (GRU) with a linear readout layer explicitly trained to reconstruct movement behaviour from neural activity. 

For all autoencoding models we use a separately trained GRU network to predict behaviour from the day 0 training session latent space. We do not include ADAN \citep{Farshchian2019AdversarialInterfaces}, NoMAD \citep{Karpowicz2022.04.06.487388} or the generative model by \citet{Wen2021RapidModelling} as all require training data from a held out session or subject to be effective. We also do not test against \citet{gonschorek2021removing} or \citep{https://doi.org/10.48550/arxiv.2202.06159} as these approaches require many training sessions to be effective in predicting behaviour from an unseen session whereas we aim to do this with just one training session.

\section{Results}

\begin{figure*}[h!]
\begin{center}
\includegraphics[width=0.72\textwidth]{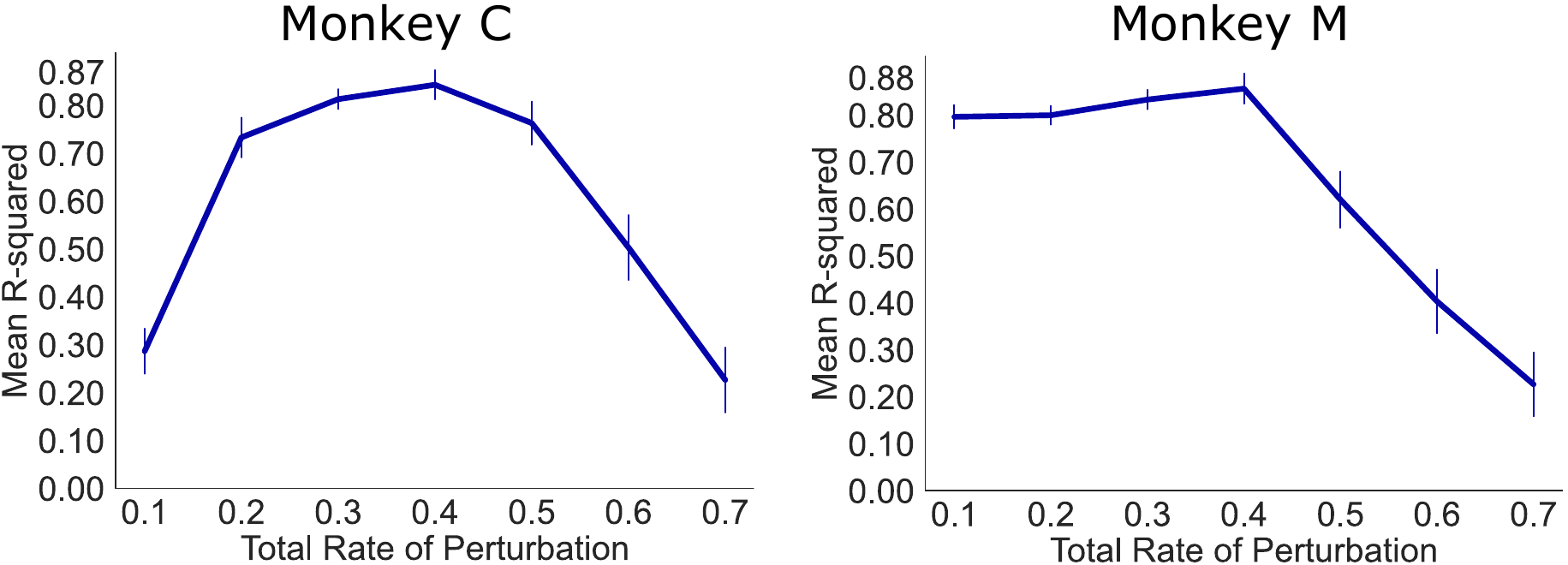}
\caption[dropout]{Behaviour decoding performance on an immediately subsequent unseen session (day 1) of CAPTIVATE at different rates of total perturbation. Total perturbation rate is the sum of the rates of perturbations A) - D) outlined in section 4, each of which are applied at equal rates.}
\label{fig:dropout}
\end{center}
\end{figure*}

Figure \ref{fig:dropout} shows behaviour decoding performance of CAPTIVATE for an unseen session that was recorded the day after the training session for different total rates of perturbation. A total perturbation rate of 40\% (i.e a rate of 10\% for each perturbation A) - D) in section 4) for both monkeys appears to be optimal. At perturbation rates above 40\%, neural activity from perturbed day 0 train trials with a particular target movement direction begin to resemble original trials of other movement directions, and thus hurt alignment. Perturbation rates below 40\%, particularly for Monkey C, are not sufficient to simulate the inter-session variability between day 0 and day 1. Training the neuron locator RNN on a total perturbation rate of 40\% for both monkeys yields 85\% and 93\% accuracy on predicting original neuron position from day 0 perturbed trials from Monkey C and Monkey M respectively. Indeed, the neuron locator network is 76\% accurate at identifying original neuron position in a simulated unseen session created with a total perturbation rate of 80\% (see Appendix D, Figure 10).

\begin{figure*}[h!]
\begin{center}
\includegraphics[width=0.80\textwidth]{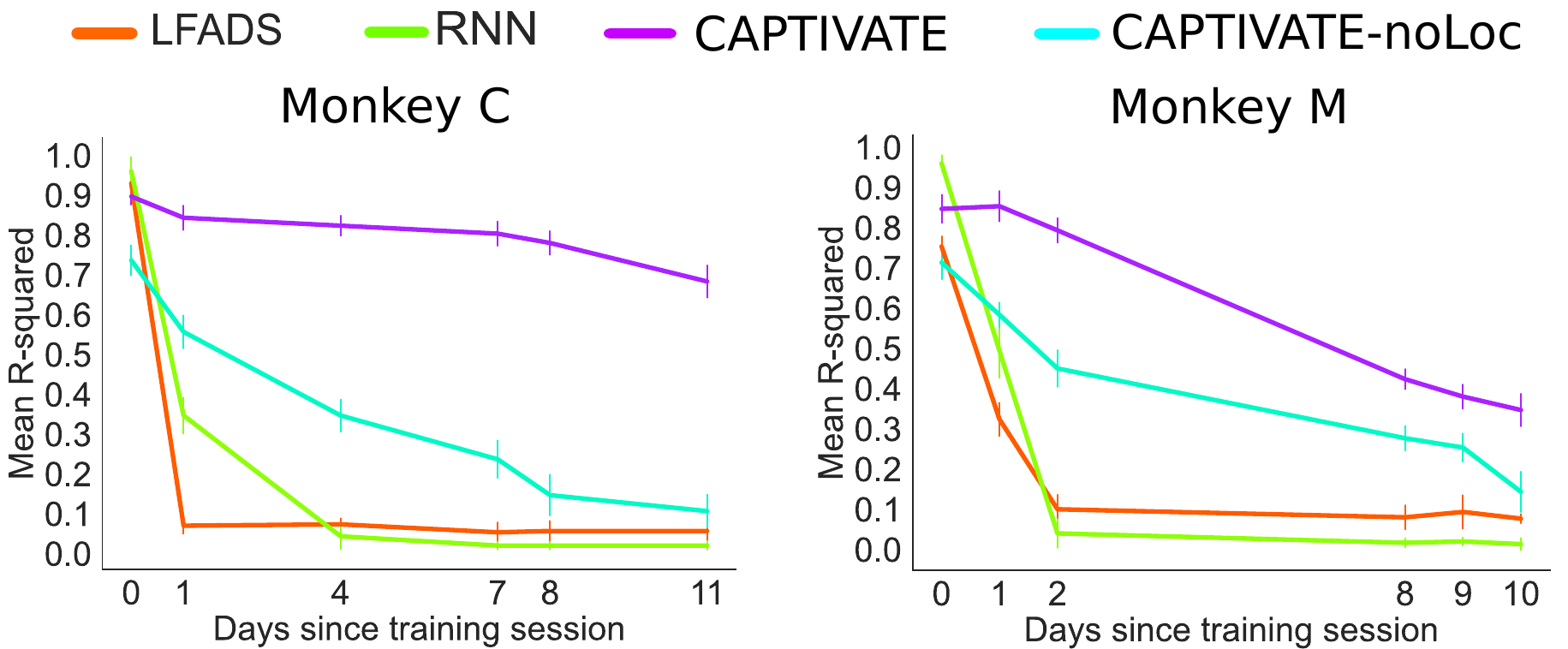}
\caption[Results]{Behaviour prediction performance when testing all models on 30\% of held-out trials from day 0 and subsequent days of completely unseen recording sessions. We report the mean $R^2$ between the inferred and true x,y positions. Each model is tested on held out trials from day 0 and trials from unseen sessions recorded an increasing number of days into the future from the original training session (day 0) for both monkeys. Each day 0 train session is run 10 times with different random seeds, with error bars showing standard deviation when applied to each unseen session.}
\label{fig:results}
\end{center}
\end{figure*}

Using the optimal rate of 40\% of perturbation to trials from both monkeys when training CAPTIVATE leads to the results summarised in Figure \ref{fig:results}. For both monkeys we see high behaviour decoding performance on the unseen session from day 1, surpassing previous methods. CAPTIVATE maintains high behaviour decoding performance for Monkey C on an unseen session up to 8 days after the day 0 training session was recorded. CAPTIVATE also accurately maps neurons from trials across unseen sessions of Monkey C up to 8 days into the future to known neurons from trials of the day 0 train session (see Appendix C, Figure 9). Notably, behaviour decoding for Monkey C is much more stable for future unseen sessions than for Monkey M. This is likely due to sessions from Monkey C containing more than 3 times as many neurons as Monkey M. However, we see in Appendix E, Figure 11 that training CAPTIVATE with 20 neurons from the Monkey C day 0 session is sufficient to achieve an $R^2$ of 0.68 when testing on 20 neurons of the day 8 session, indicating our model can be robust to a low number of neurons.  

Notably in the case of Monkey M, day 1 decoding performance is high at all levels of perturbation from 0.1 to 0.4 (Figure \ref{fig:dropout}), therefore it is likely that the session to session variability between day 0 and day 1 is small. Thus, for a subject with fewer neurons in recorded data, CAPTIVATE may only require a low rate of total perturbation when aligning nearby unseen sessions.

\begin{figure}[h!]
\begin{center}
\includegraphics[width=0.875\textwidth]{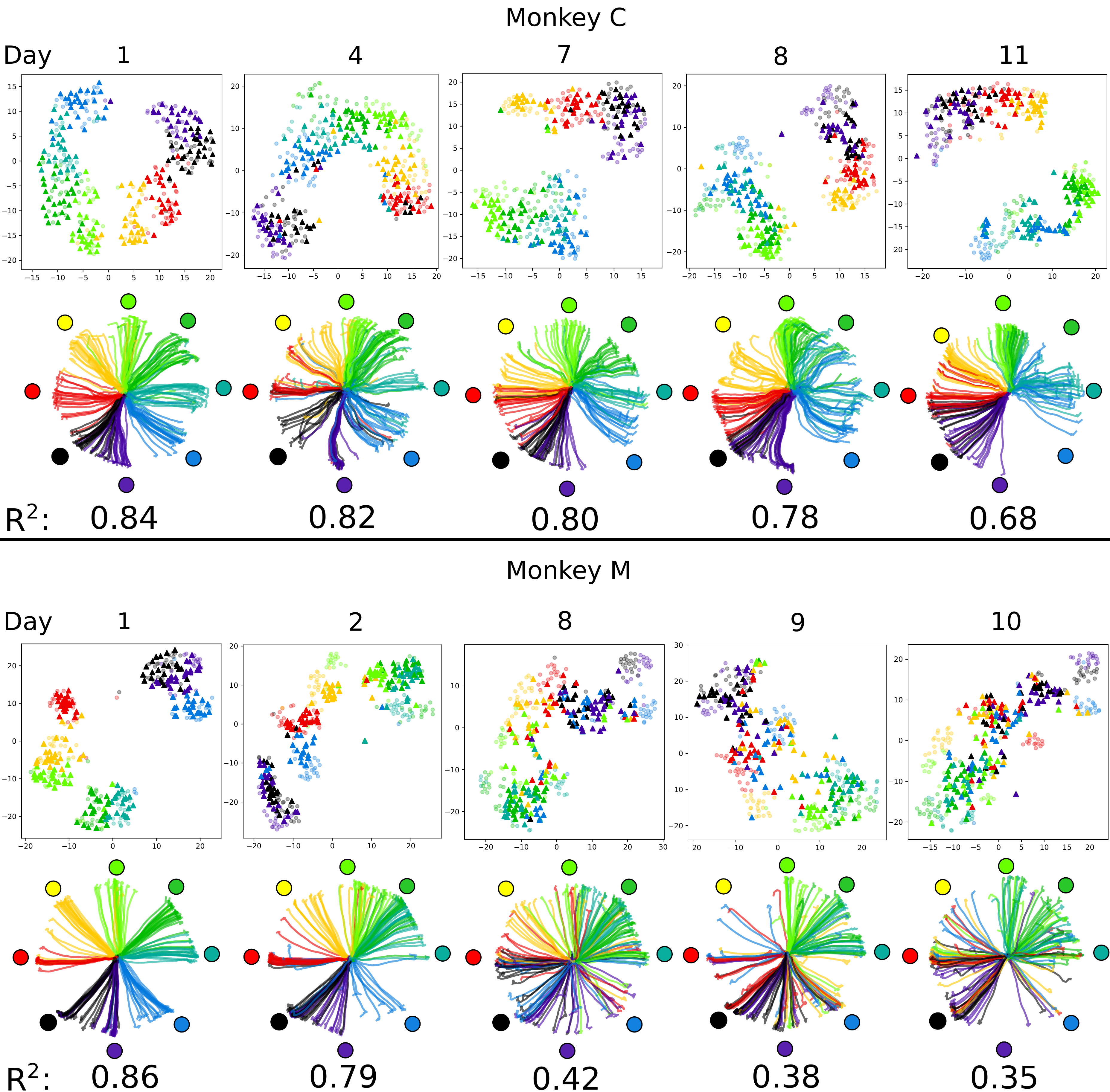}
\caption[behaviour]{For each monkey, \textit{Top row}: t-SNE embeddings of latent space for CAPTIVATE when applied to each unseen session. In each embedding, points denoted by a circle are trials from the day 0 training session. Points denoted by a triangle are trials from the named unseen session. Each colour represents a target direction for the centre-out reach task. \textit{Bottom row}: Predicted 2D monkey hand position of trials using a separately trained RNN decoder trained only on the day 0 latent space of CAPTIVATE when applied to each unseen session, with mean $R^2$ between all positions of each predicted and ground truth trajectory shown across all trials in a given unseen session.}
\label{fig:behaviour}
\end{center}
\end{figure}

CAPTIVATE-noLoc, Vanilla LFADS or an RNN model cannot capture session-to-session variability even for the day 1 unseen session, as shown in Figure \ref{fig:behaviour}. CAPTIVATE-noLoc cannot accurately reconstruct original trials from perturbed variations of the day 0 train session, but has a similar day 0 and day 1 session behaviour decoding accuracy, implying our perturbations closely mirror inter-session variability. This indicates that poor performance of CAPTIVATE-noLoc on both monkeys is due to the inability of the encoder of this model to recognise known neurons and thus, shows how crucial the neuron locator network is in recognising known neuron ensembles in unseen recordings. 

LFADS is trained solely on unperturbed trials and so cannot recognise the shifts that occur between sessions as variations of the day 0 training session, and thus cannot create an appropriate latent encoding. The RNN model is also trained on unperturbed trials and is even less robust to later unseen sessions than LFADS, however, this RNN baseline can recognise some behaviour in both monkeys for the day 1 unseen session, indicating a relatively low level of variability in adjacent day recordings. Similarly, we see a mean $R^2$ of 0.37 ($\pm$ 0.02) when training an RNN on day 7 for monkey C and testing on day 8 and an $R^2$ of 0.42 ($\pm$ 0.04) when training an RNN on day 9 and testing on day 10 for monkey M. Importantly, none of these models overfit as they yield high decoding accuracies for a held-out portion of day 0 trials for both monkeys and for all models, especially the RNN. Therefore the performance drop of the RNN model when applied to unseen sessions is a clear indication of substantial variations between sessions.

Figure \ref{fig:behaviour} shows t-SNE visualisations of the latent space and behaviour predictions made from the latent space of CAPTIVATE when trained on the day 0 session and applied to unseen sessions. For Monkey C, the majority of trials from all unseen sessions are correctly aligned with the corresponding trials in the training data set (Figure \ref{fig:behaviour}, compare dots and triangles in the t-SNE plots where colour indicates movement direction; note that the latent space is well partitioned by behaviour although the model is only trained on neural activity). Occurrences where the unseen trials correctly overlap known train trials, in turn, yields correctly decoded behaviour. The alignment becomes progressively worse for later sessions, and as the alignment is less precise, behaviour predictions also become worse. In contrast, for Monkey M the alignment of trials beyond day 2 becomes increasingly worse, a consequence of the smaller number of neurons in the recording.


\begin{table}[h!]

\caption{Mean decoding performance effects of ablating individual perturbations when training on day 0 session and tested on immediately subsequent (day 1) unseen session for both monkeys. For reference, the full CAPTIVATE model trained on day 0 achieves mean decoding $R^2$ performance of 0.84 ($\pm$ 0.02) on monkey C and 0.86 ($\pm$ 0.03) on monkey M when applied to the day 1 session.}
\label{tab:ablations}
\begin{center}
\begin{tabular}{ c | c c c c c c} 
\hline
 Ablation & No-Replace & No-Add & No-Remove & No-Delete & No-Jitter & No-Reorder\\ 
 \hline
C Mean $R^2$ & 0.66 & 0.79 & 0.74 & 0.49 & 0.77 & 0.70\\
& ($\pm$ 0.03) & ($\pm$ 0.01) & ($\pm$ 0.02)& ($\pm$ 0.04)& ($\pm$ 0.01)& ($\pm$ 0.03)\\
\hline

M Mean $R^2$ & 0.71 & 0.75 & 0.81 & 0.63 & 0.81 & 0.77\\ 
& ($\pm$ 0.03) & ($\pm$ 0.02) & ($\pm$ 0.01)& ($\pm$ 0.03)& ($\pm$ 0.01)& ($\pm$ 0.02)\\
\hline
\end{tabular}
\end{center}

\end{table}

Ablations of individual perturbations (as outlined in Figure \ref{fig:perturbations}) applied when training on the day 0 session reveal that perturbations which introduce randomly generated neurons and alter the continuous ordering of neurons have the highest impact on unseen session behaviour decoding performance. 
This analysis is summarised in Table \ref{tab:ablations} and shows that neuron deletions, replacements and probe shifts cause the majority of inter-session neuron ensemble variability. Nonetheless, a combination of all perturbations are necessary for the decoding performance achieved by CAPTIVATE in Figure \ref{fig:results}. We also train CAPTIVATE without the alignment loss in Eq. \ref{eq:align}, which produces a behaviour decoding mean $R^2$ of 0.82 on Monkey C and 0.85 on Monkey M on trials from the day 1 unseen session. This minimal drop in decoding performance when training without an explicit alignment loss is consistent with results from \citep{liu2021drop}. Additionally, decoding performance across unseen sessions when training on day 0 and day 11 sessions separately is almost symmetrical (as shown in Appendix F, Figure 12), indicating that our model can effectively capture neural variability from unseen sessions both forwards and backwards in time. We further assess robustness by testing CAPTIVATE on a variable number of neurons across sessions (similar to a real BCI setting) and show good generalisation, even surpassing performance of the model trained with 55 neurons (as in Figure \ref{fig:results}) for some unseen sessions (see Appendix G, Figure 13).

\section{Discussion}
In this paper we use a self-supervised approach, CAPTIVATE, to train a model to recognise and correct for session-to-session variability in neural recordings. We then show that the combination of this approach with a latent variable model that identifies low-dimensional dynamics in neural activity yields a model that is now robust variability between recordings sessions. The model is capable of successfully predicting behaviour with high accuracy from unseen sessions, surpassing previous work by \cite{https://doi.org/10.48550/arxiv.2202.06159} when comparing against subsequent day decoding performance. Furthermore, our approach leads to relatively high and stable behaviour decoding performance on unseen sessions many days into the future when a sufficient number of neurons are persistent across sessions. As a result, this method performs better for data sets with more recorded neurons (Monkey C), while for fewer neurons the performance degrades more quickly, only producing good results for sessions close in time to the training session (Monkey M).


With CAPTIVATE we achieve stable behaviour decoding performance for up to 8 days, which is followed by a slow decline in performance. The decline is due to an increase in variability that could no longer be compensated. This would require a model to correct even stronger perturbations, but training a model this way leads to an overall decrease in performance even for short time intervals (Figure \ref{fig:dropout}). Therefore long-term stable decoding currently still requires re-training of the components of a latent variable encoder model such that the altered neural dynamics are re-aligned with the latent dynamics \citep{Wen2021RapidModelling,Karpowicz2022.04.06.487388, Farshchian2019AdversarialInterfaces}. Equally, our model fails to successfully decode behaviour from recordings from an unseen animal (not illustrated) as this requires a more complex mapping function between activity and latent space \citep{Wen2021RapidModelling}.

\bibliography{biblio,references}
\bibliographystyle{abbrvnat}
\section*{Checklist}

\begin{enumerate}

\item For all authors...
\begin{enumerate}
  \item Do the main claims made in the abstract and introduction accurately reflect the paper's contributions and scope?
    \answerYes{See Results, Section 6.}
  \item Did you describe the limitations of your work?
    \answerYes{See Discussion, Section 7.}
  \item Did you discuss any potential negative societal impacts of your work?
    \answerNA{}
  \item Have you read the ethics review guidelines and ensured that your paper conforms to them?
    \answerYes{}
\end{enumerate}

\item If you are including theoretical results...
\begin{enumerate}
  \item Did you state the full set of assumptions of all theoretical results?
    \answerNA{}
        \item Did you include complete proofs of all theoretical results?
    \answerNA{}
\end{enumerate}

\item If you ran experiments...
\begin{enumerate}
  \item Did you include the code, data, and instructions needed to reproduce the main experimental results (either in the supplemental material or as a URL)?
    \answerYes{Perturbation and model instructions in sections 4 and 5 respectively. Code included in Supplement.}
  \item Did you specify all the training details (e.g., data splits, hyperparameters, how they were chosen)?
    \answerYes{In Section 5 and further implementation details in Appendix A.}
        \item Did you report error bars (e.g., with respect to the random seed after running experiments multiple times)?
    \answerYes{See figures in Section 6.}
        \item Did you include the total amount of compute and the type of resources used (e.g., type of GPUs, internal cluster, or cloud provider)?
    \answerYes{Yes, see details of compute used in Section 7.}
\end{enumerate}

\item If you are using existing assets (e.g., code, data, models) or curating/releasing new assets...
\begin{enumerate}
  \item If your work uses existing assets, did you cite the creators?
    \answerYes{See Section 3.}
  \item Did you mention the license of the assets?
    \answerNA{}
  \item Did you include any new assets either in the supplemental material or as a URL?
    \answerYes{Code included in supplemental material.}
  \item Did you discuss whether and how consent was obtained from people whose data you're using/curating?
    \answerNA{}
  \item Did you discuss whether the data you are using/curating contains personally identifiable information or offensive content?
    \answerNA{}
\end{enumerate}

\item If you used crowdsourcing or conducted research with human subjects...
\begin{enumerate}
  \item Did you include the full text of instructions given to participants and screenshots, if applicable?
    \answerNA{}
  \item Did you describe any potential participant risks, with links to Institutional Review Board (IRB) approvals, if applicable?
    \answerNA{}
  \item Did you include the estimated hourly wage paid to participants and the total amount spent on participant compensation?
    \answerNA{}
\end{enumerate}

\end{enumerate}


\newpage
\appendix

\section{Implementation and training details}
\label{app:modeldetails}
Below are implementation details for the CAPTIVATE model.
\begin{center}
\begin{tabular}{ |p{6cm}|p{1.7cm}|p{4.6cm}|  }

\hline
\multicolumn{3}{|c|}{CAPTIVATE} \\
\hline
Parameter & Value & Notes \\
\hline
Neuron Locator Network &  & Layer Normalisation on all layers\\
\quad- RNN Units & 784 X 3   & Stacked Gated Recurrent Unit \\
\quad- $W_{pos}$ Units & 1024 X 3 & Non-linear layer \\
\quad- $W_{pos}$ Dropout & 0.5 &  \\
\quad- $W_{pos}$ L2 Regularisation & 100.0 &  \\
\hline
Sequential Autoencoder Encoder &  & \\
\quad- RNN Units & 784 X 3   & Stacked Gated Recurrent Unit \\
\quad - RNN L2 Kernel Regularisation & 0.1 &  \\
\quad- RNN L2 Recurrent Regularisation & 0.1 & \\
\quad- $W_{enc}$ Units & 1024 X 3 & Non-linear layer \\
\quad- $W_{enc}$ L2 Regularisation & 0.1 &  \\
\quad- Latent space dimension & 64 &  \\
\hline
Sequential Autoencoder Generator &  & \\
\quad- RNN Units & 512 X 3  &  Stacked Gated Recurrent Unit \\
\quad - RNN L2 Kernel Regularisation & 1.0 &  \\
\quad- RNN L2 Recurrent Regularisation & 1.0 & \\
\quad- $W_{fac}$ Units & 512 & Non-linear layer \\
\hline
Training &  & \\
\quad- KL divergence weighting ($\lambda_{kl}$)   & 0.02 to 1.0 & Rising exponentially  \\
\quad- Batch size (Train Neuron Locator)   &  16 &   \\
\quad- Batch size (Train Seq. Autoencoder)   &  4 &   \\
\quad- Learning rate (Train Neuron Locator)   &  0.0001 &  Adam Optimizer\\
\quad- Learning rate (Train Seq. Autoencoder)   &  0.00001 &  Adam Optimizer \\
\hline

\end{tabular}
\end{center}
\newpage

\section{Changes in recorded neural activity across sessions}
\begin{figure*}[h!]
\begin{center}
\includegraphics[width=1.\textwidth]{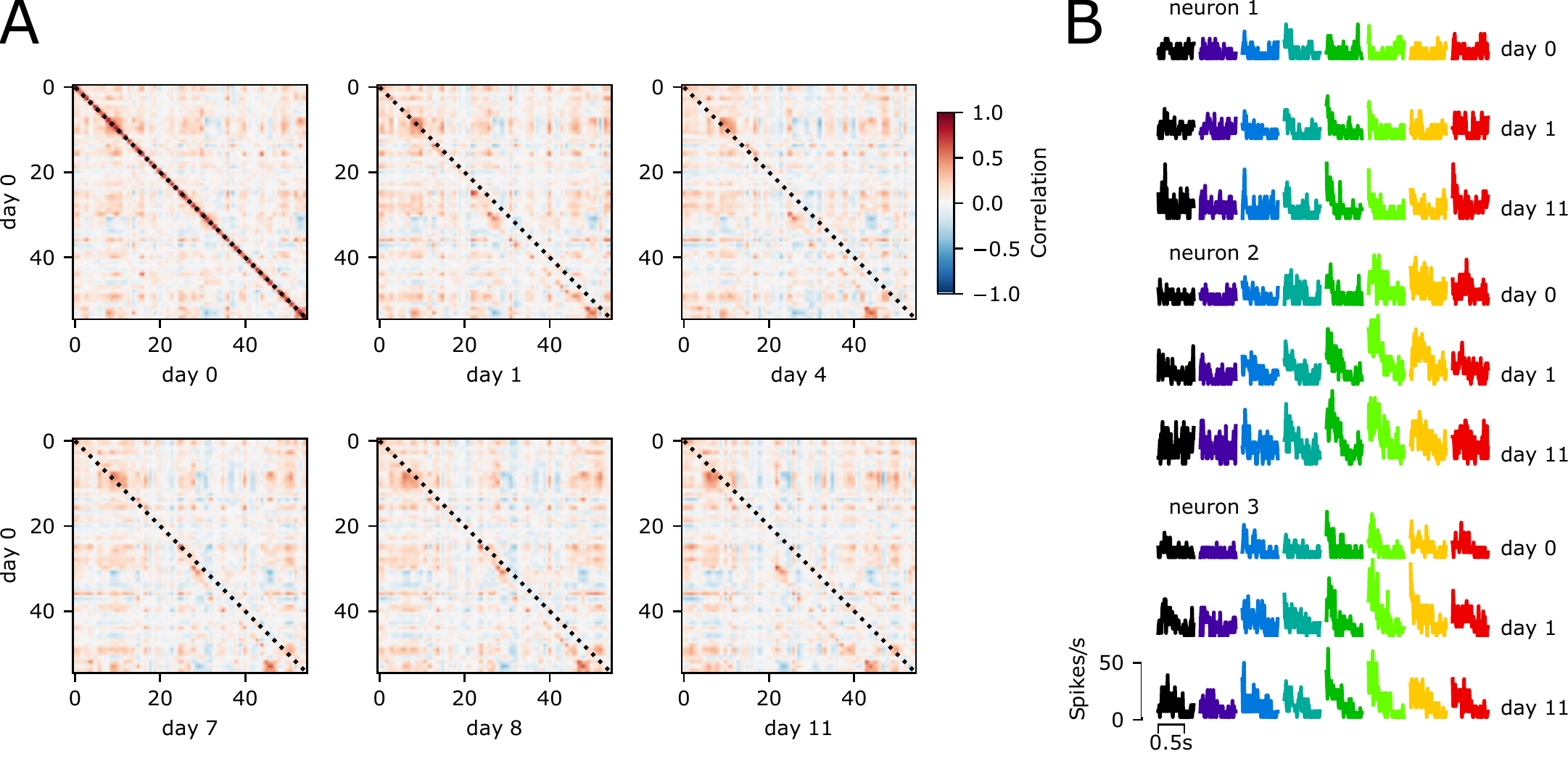}
\caption[changes]{\textbf{A}, Pairwise correlations of trial-averaged activity of single neurons between two sessions. For each neuron, the average firing rate was computed for each of the eight movement directions (see part B for examples) at 10ms resolution. The activity for the eight movement directions was concatenated and the Pearson correlation coefficients computed between all neuron pairs. Each plot shows the correlation matrix for activity from session from a different day and activity from the first day (day zero, the training data set in Figure \ref{fig:results}). This analysis shows that some neurons from the first session can be matched to neurons recorded at subsequent days, but the relative position of these matched neurons in the recording tends to shift (see high off-diagonal correlations). As the average correlations do not change systematically over this period of time (not illustrated), the gradual changes in neuron identity is a main factor that prevents reliable decoding from unseen sessions in previous models. \textbf{B}, Examples of trial-averaged firing rates of three neurons that were tracked over all recording sessions. This matching is based on the similarity of the firing rates, experimentally it is hard to determine if these are indeed the same neurons. In all cases, the time course of the activity is similar and shows consistent differences between trial type (indicated by colour) across sessions. Also note that while these neurons appear to reliably encode movement direction, the activity of a single neuron  alone is too noisy to allow for reliable direction decoding from single trials, instead a population decoding approach is required. All data illustrated here is from Monkey C.  }
\label{fig:changes}
\end{center}
\end{figure*}

\newpage

\section{CAPTIVATE accurately maps perturbed neurons and neurons from unseen sessions to known neurons from the Day 0 training session}
CAPTIVATE is trained by mapping perturbed trials to known trials. If trials from unseen sessions are similar to the perturbed trials then generalisation to these sessions is possible. Therefore, we aim for the encoder network of CAPTIVATE to map perturbed trials and trials from unseen sessions to day 0 trials. This entails that neurons across unseen sessions (even after neural drift and ensemble change) are mapped directly to neuron positions of the day 0 session at the session. For trials of each movement direction from unseen sessions, we expect that the trial average firing rates of these neurons will map to the day 0 average firing rates for each neuron. As seen below for 4 neurons across 3 sessions (2 unseen), the CAPTIVATE generator network produces trial average firing rates matching the day 0 train session firing rates.
\begin{figure*}[h!]
\begin{center}
\includegraphics[width=0.63\textwidth]{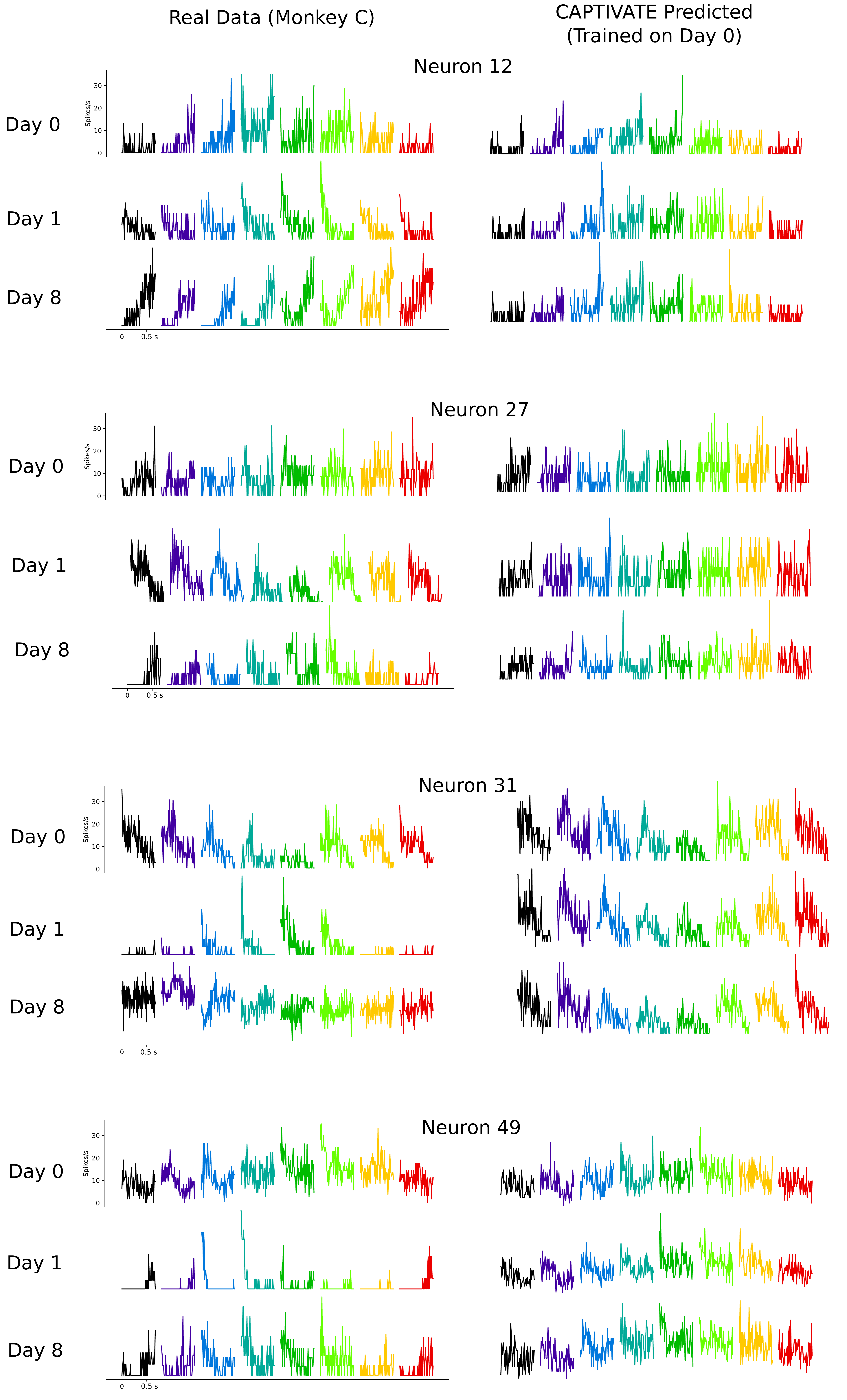}
\caption[chewieneuronsrates]{CAPTIVATE is trained on the day 0 session of Monkey C. On the left we show real trial averaged firing rates for each movement direction across 4 randomly selected neurons across the day 0 session and 2 unseen sessions. On the right we show predictions from the generator network of CAPTIVATE. If generalisation is achieved the generator should accurately map neurons across unseen sessions to the neurons of day 0. We see that this is the case as the predicted firing rates are closely matched in the unseen sessions to the day 0 firing rates across movement directions.}
\label{fig:chewie_neurons_rates}
\end{center}
\end{figure*}

\newpage

\section{Neuron Locator performance over simulated neural variation}
As we do not have ground truth neuron identities from unseen sessions (with respect to the day 0 train session), we simulate inter-session variability by increasing perturbation rate and testing against CAPTIVATE trained on the day 0 session from monkey C with a total perturbation rate of 0.4 (as in the results shown in Figure 6). We see that the neuron locator network of CAPTIVATE can predict neuron identity with 68\% accuracy even at a very high total perturbation rate of 1.0. 
\begin{figure*}[h!]
\begin{center}
\includegraphics[width=0.71\textwidth]{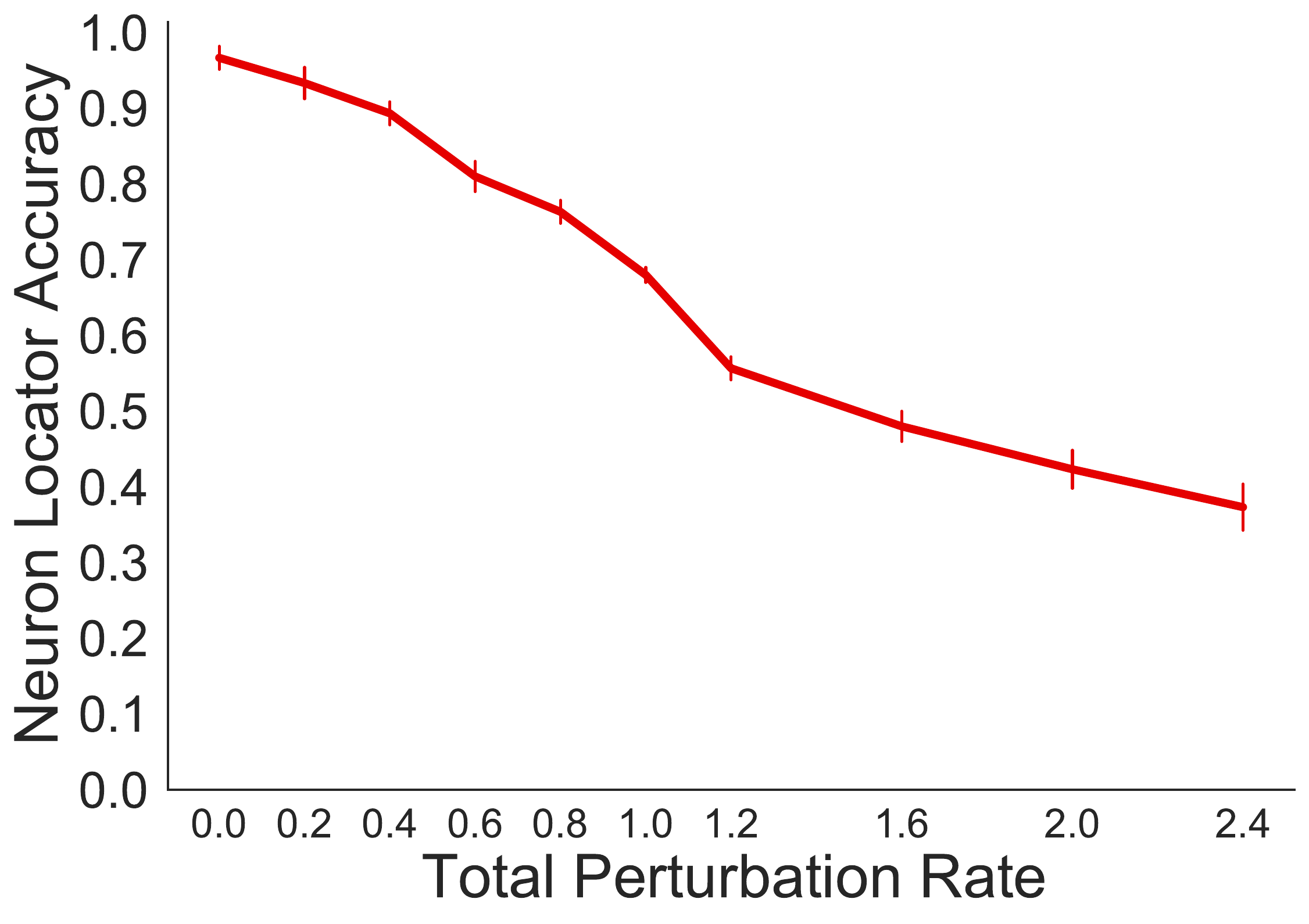}
\caption[neuronacc]{Neuron locator network accuracy when predicting neuron identity (with respect to unperturbed day 0 monkey C train session) as the total rate of perturbation is increased. We are simulating neural drift and ensemble shift across sessions. As we know the ground truth neuron identities, we can assess how well the neuron locator can predict neuron identity.}
\label{fig:neuron_acc}
\end{center}
\end{figure*}

\newpage

\section{Training and testing CAPTIVATE with different numbers of original neurons}

Here we test the varying numbers of neurons across sessions of Monkey C when using CAPTIVATE. We see that only 20 neurons are required across sessions for good generalisation for up to 8 days.
\begin{figure*}[h!]
\begin{center}
\includegraphics[width=0.71\textwidth]{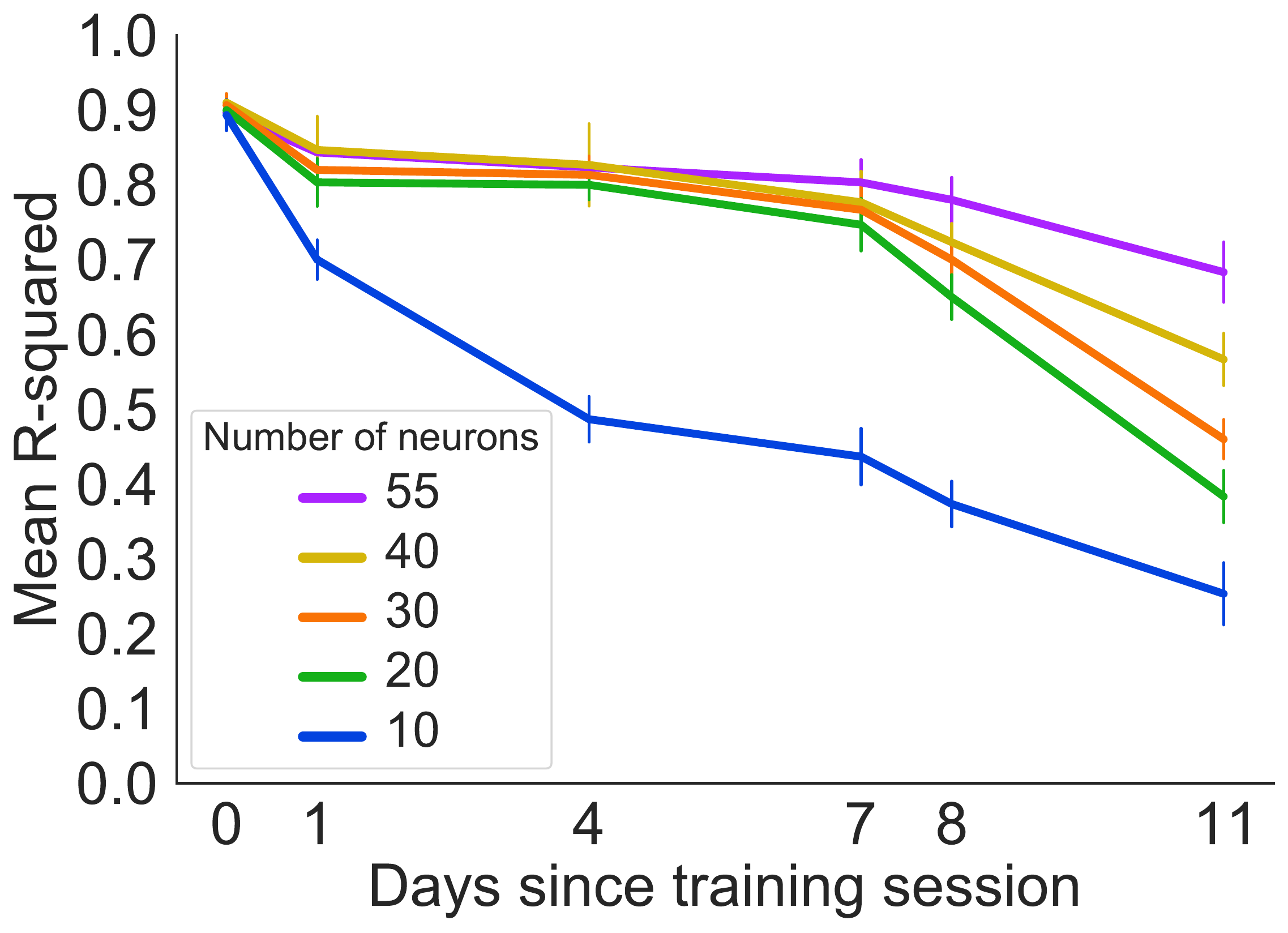}
\caption[reducingneurons]{Behaviour prediction performance when training CAPTIVATE on varying numbers of neurons of the day 0 session recorded from Monkey C and testing on all other unseen sessions of monkey C, using the same number of neurons as used in the training session. We also test all neuron number variations of CAPTIVATE on a held out portion of trials from day 0. We report the mean $R^2$ between the inferred and true x,y positions for the entire movement trajectory of each trial. Each day 0 train session is run 10 times with different random seeds, with error bars showing standard deviation when applied to each unseen session.}
\label{fig:reducingneurons}
\end{center}
\end{figure*}

\newpage
\section{Changing calibration session}
Here we show that by training our model on perturbed trials we can generalise to neural drift and recording array movement. CAPTIVATE accounts not only for session variability in the future but also in the past.
\begin{figure*}[h!]
\begin{center}
\includegraphics[width=0.71\textwidth]{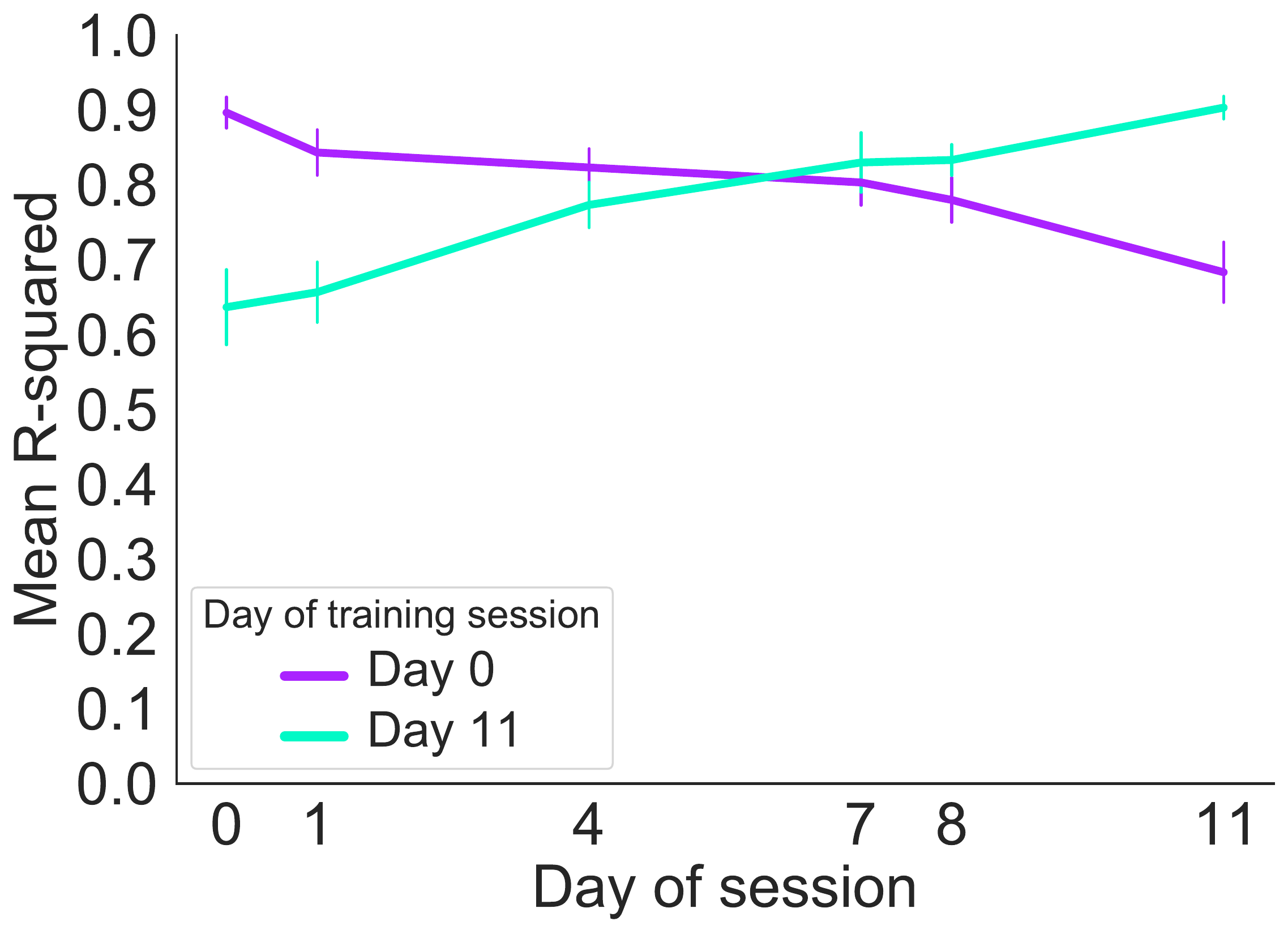}
\caption[trainlast]{Behaviour prediction performance when separately training CAPTIVATE on the day 0 and day 11 sessions of Monkey C and testing on all other unseen sessions. We see that performance across unseen sessions when training on these sessions is almost symmetrical, indicating that our model can effectively capture neural variability from sessions both backwards and forwards in time.}
\label{fig:train_last}
\end{center}
\end{figure*}

\newpage
\section{Variable neuron number per session}
When using an implanted recording array we may lose electrodes or neurons due to spike sorting error over a period of time. Here we show our model can account for this variable neuron number.
\begin{figure*}[h!]
\begin{center}
\includegraphics[width=0.71\textwidth]{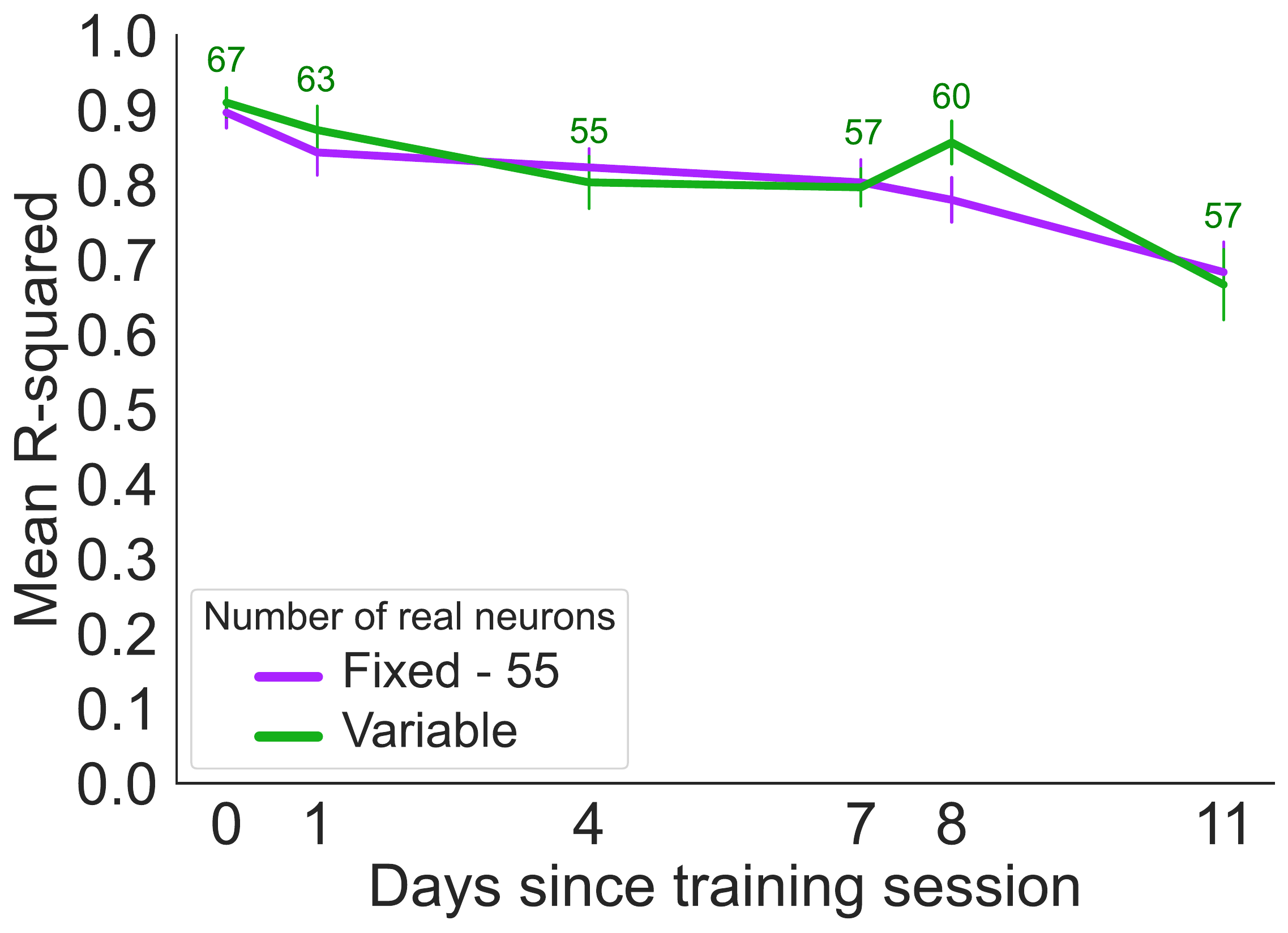}
\caption[trainvariable]{We test CAPTIVATE with a variable number of neurons per session of recording from Monkey C. In our original experiment we only utilise the first 55 neurons of each recording session as this is the minimum number across all sessions. Here we use every neuron available per session (number of neurons per session shown in Figure) and train CAPTIVATE on the day 0 session with 67 neurons. For all other sessions we add randomly generated neurons to compensate. We see that CAPTIVATE is robust to the number of original neurons being variable across sessions. Note the increase in generalisation performance when the model is applied to the day 8 session. This is due to this session having a relatively high number (60) of original neurons, and is thus easier for the model to map trials from this session to known trials from the day 0 training session than it is from other later unseen sessions.}
\label{fig:train_variable}
\end{center}
\end{figure*}

\newpage
\section{Testing trained model on known neural variability}
We test the resilience of our whole model against an increasing total rate of perturbation in order to ascertain how much variability the model can account for. For the results below, CAPTIVATE is trained with a total perturbation rate of 0.4 on the day 0 session of Monkey C. Note that the model is trained to map perturbed trials to original unperturbed trials.
\begin{figure*}[h!]
\begin{center}
\includegraphics[width=0.99\textwidth]{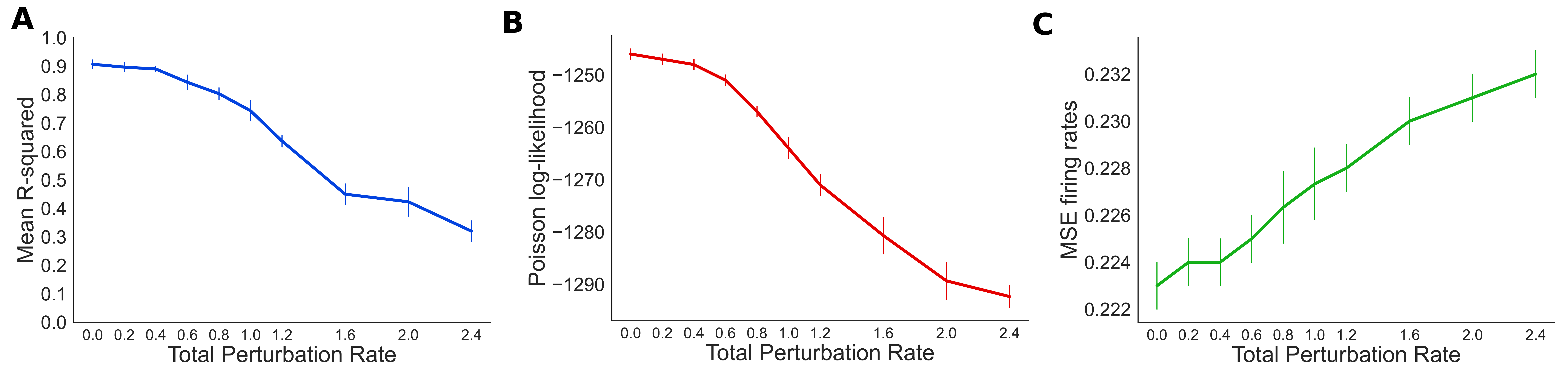}
\caption[simulation]{We train CAPTIVATE on trials from the day 0 session of Monkey C with a 0.4 total rate of perturbation on each trial. We then test the trained model on trials of the same session but with increasing rates of total perturbation applied to trials. A) Mean r-squared error of movement predicted from the latent space of the model vs. real movement trajectory of each trial. B) Mean Poisson log-likelihood for neural activity reconstruction by the model generator of original day 0 unperturbed trials. C) Mean squared error of model predicted firing rates vs. real firing rates of original unperturbed day 0 trials.}
\label{fig:simulation_plots}
\end{center}
\end{figure*}

\newpage
\section{Testing trained model on known neural variability across held-out trials}
We test the resilience of our whole model against an increasing total rate of perturbation in order to ascertain how much variability the model can account for. For the results below, CAPTIVATE is trained with a total perturbation rate of 0.4 on 70\% of the trials of the day 0 session of Monkey C. We show test performance on 30\% of the trials of the day 0 session which are withheld from training. Note that the model is trained to map perturbed trials to original unperturbed trials.
\begin{figure*}[h!]
\begin{center}
\includegraphics[width=0.99\textwidth]{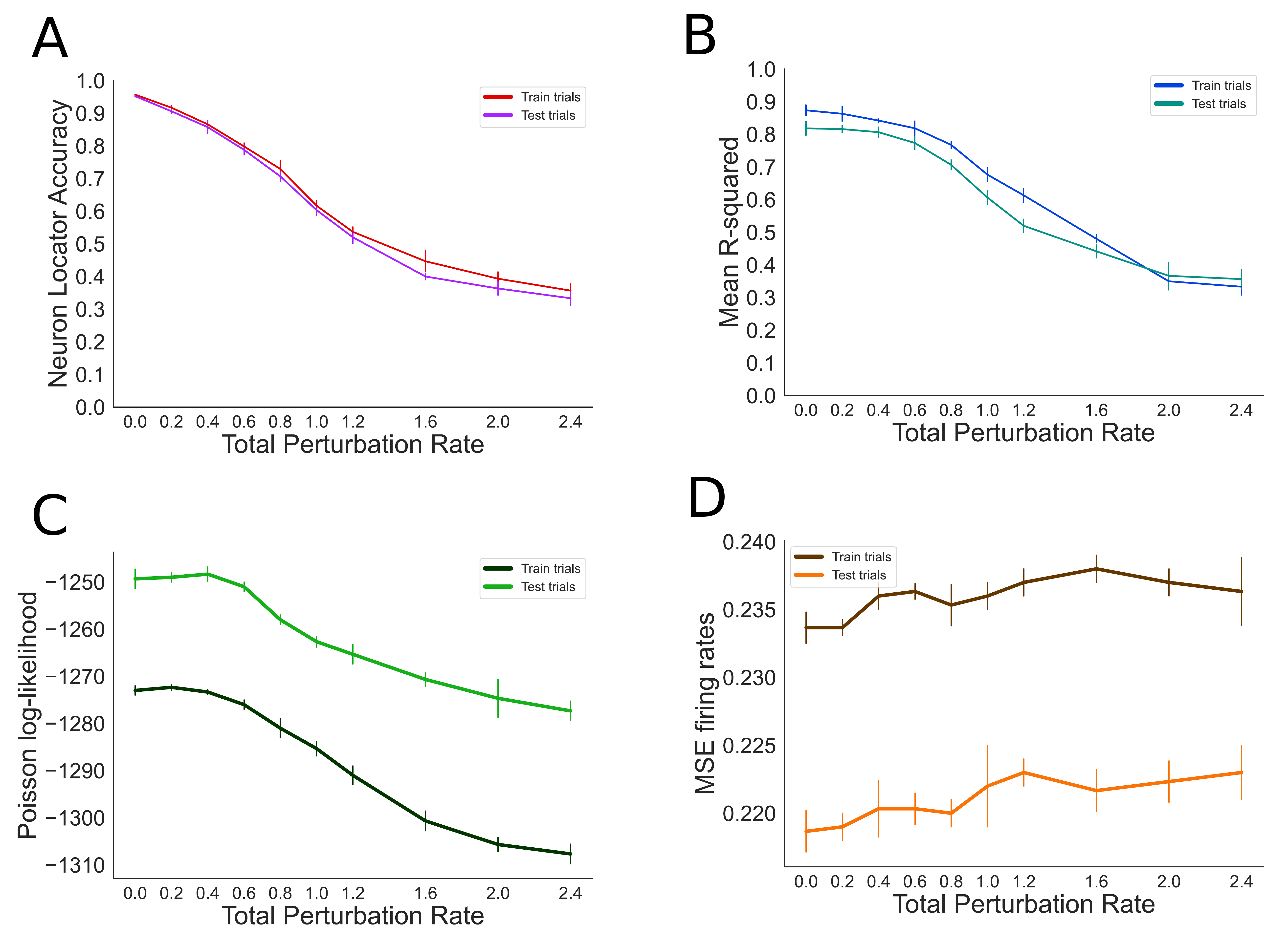}
\caption[simulation70]{We train CAPTIVATE on 70\% of the trials from the day 0 session of Monkey C with a 0.4 total rate of perturbation on each trial. We then test the trained model on the remaining 30\% of trials of the same session but with increasing rates of total perturbation applied to these held-out trials. A) Neuron locator network accuracy when predicting neuron identity (with respect to unperturbed day 0 Monkey C train session) as the total rate of perturbation is increased. We are simulating neural drift and ensemble shift across sessions. As we know the ground truth neuron identities, we can assess how well the neuron locator can predict neuron identity. B) Mean r-squared error of movement predicted from the latent space of the model vs. real movement trajectory of each trial. C) Mean Poisson log-likelihood for neural activity reconstruction by the model generator of original day 0 unperturbed trials. D) Mean squared error of model predicted firing rates vs. real firing rates of original unperturbed day 0 trials.}
\label{fig:simulation_plots70}
\end{center}
\end{figure*}

\end{document}